\documentclass[journal = jpclcd, manuscript = letter, layout = traditional]{achemso}

\usepackage{graphicx}
\usepackage{dcolumn}
\usepackage[ansinew]{inputenc}
\usepackage{array}
\usepackage{color}
\usepackage{amsmath}
\usepackage{amsxtra}
\usepackage{amstext}
\usepackage{amssymb}
\usepackage{latexsym}

\title{Ultrafast dynamics of excited electronic states in nitrobenzene measured by ultrafast transient polarization spectroscopy }

\author{Richard Thurston}
\affiliation{Chemical Sciences Division, Lawrence Berkeley National Laboratory, Berkeley, CA 94720 USA}

\author{Matthew M. Brister}
\affiliation{Chemical Sciences Division, Lawrence Berkeley National Laboratory, Berkeley, CA 94720 USA}

\author{Liang Z. Tan}
\affiliation{Molecular Foundry, Lawrence Berkeley National Laboratory, Berkeley, CA 94720 USA}

\author{Elio G. Champenois}
\affiliation{Chemical Sciences Division, Lawrence Berkeley National Laboratory, Berkeley, CA 94720 USA}
\alsoaffiliation{Graduate Group in Applied Science and Technology, University of California, Berkeley, CA 94720}

\author{Said Bakhti}
\affiliation{Chemical Sciences Division, Lawrence Berkeley National Laboratory, Berkeley, CA 94720 USA}

\author{Pavan Muddukrishna}
\affiliation{Chemical Sciences Division, Lawrence Berkeley National Laboratory, Berkeley, CA 94720 USA}

\author{Thorsten Weber}
\affiliation{Chemical Sciences Division, Lawrence Berkeley National Laboratory, Berkeley, CA 94720 USA}

\author{Ali Belkacem}
\affiliation{Chemical Sciences Division, Lawrence Berkeley National Laboratory, Berkeley, CA 94720 USA}

\author{Daniel S. Slaughter}
\email{dsslaughter@lbl.gov}
\affiliation{Chemical Sciences Division, Lawrence Berkeley National Laboratory, Berkeley, CA 94720 USA}

\author{Niranjan Shivaram}
\email{niranjan@purdue.edu}
\affiliation{Chemical Sciences Division, Lawrence Berkeley National Laboratory, Berkeley, CA 94720 USA}
\altaffiliation{Current address: Department of Physics and Astronomy, Purdue University, West Lafayette, IN 47907 USA}

\begin{document}

\begin{tocentry}
%
%
%
%
%

\includegraphics[width=1 \textwidth]{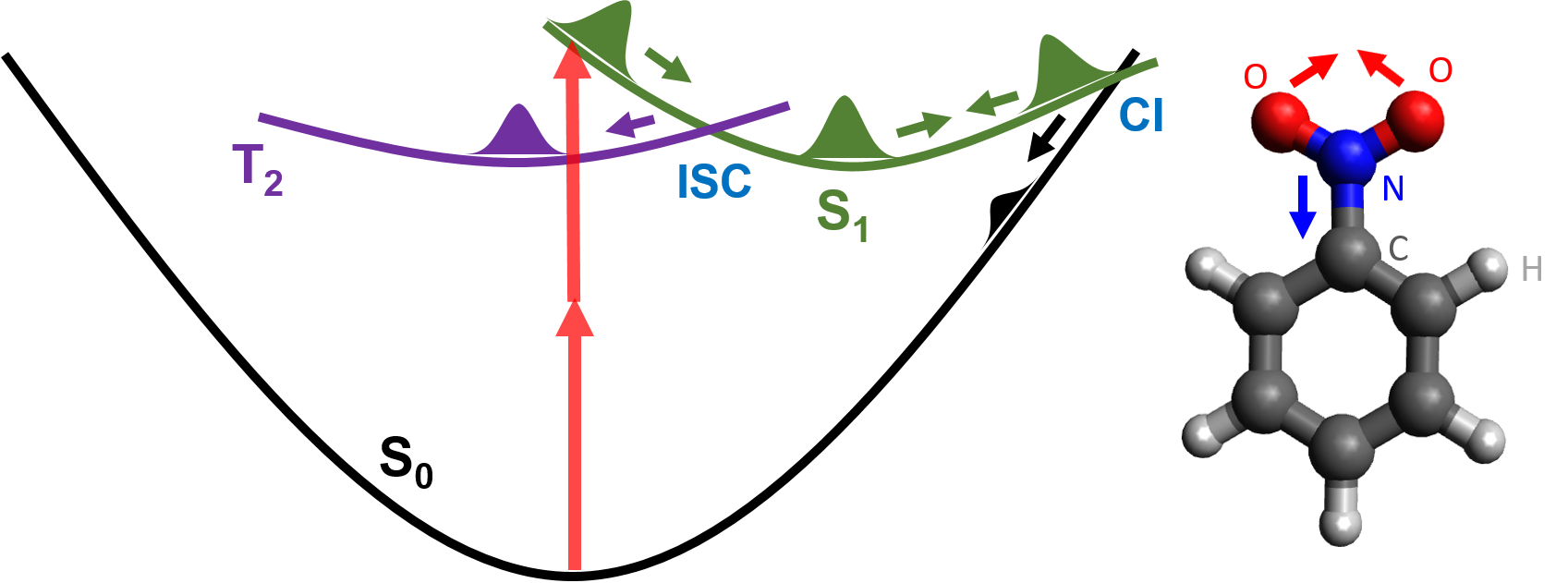}
\includegraphics[width=0.78 \textwidth]{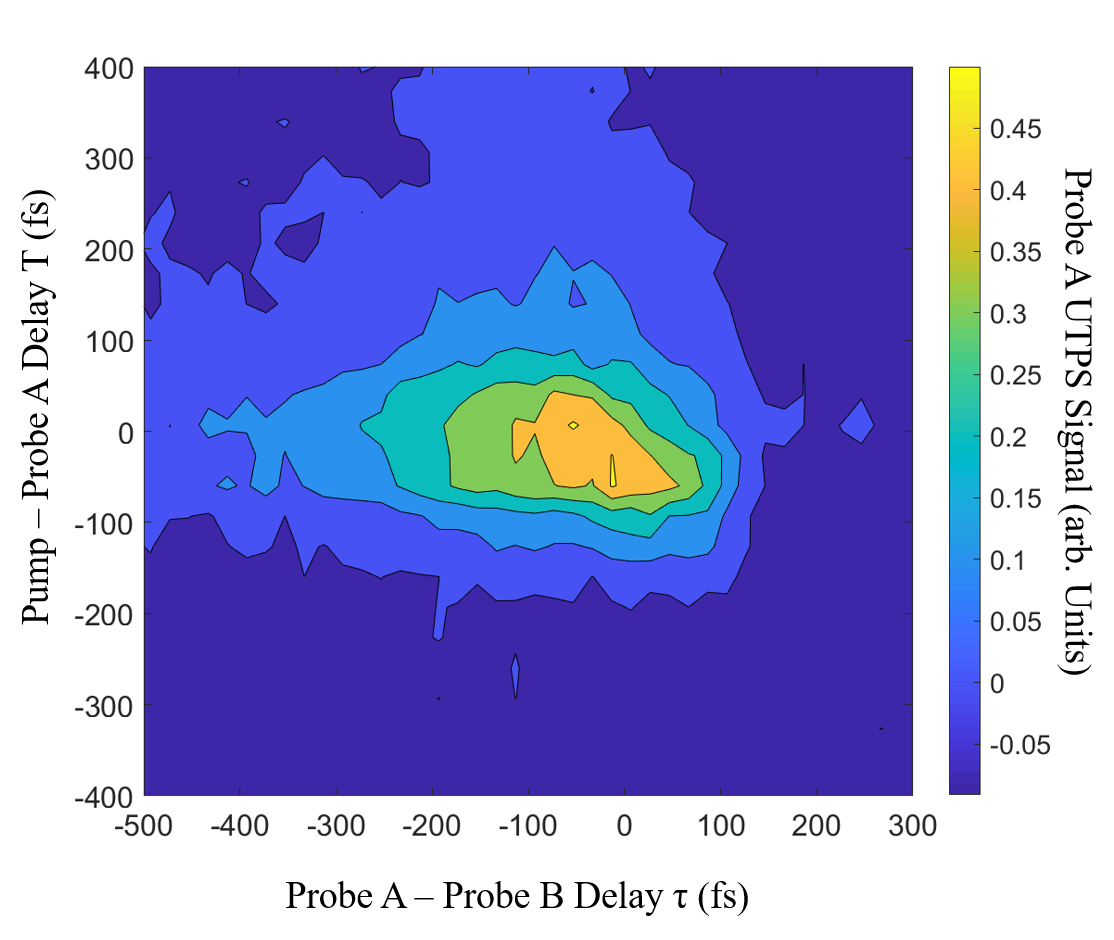}
\end{tocentry}

\begin{abstract}
We investigate ultrafast dynamics of the lowest singlet excited electronic state in liquid nitrobenzene using Ultrafast Transient Polarization Spectroscopy (UTPS), extending the well known technique of Optical Kerr-Effect (OKE) spectroscopy to excited electronic states. The third-order non-linear response of the excited molecular ensemble is highly sensitive to details of excited state character and geometries and is measured using two femtosecond pulses following a third femtosecond pulse that populates the S$_1$ excited state. By measuring this response as a function of time delays between the three pulses involved, we extract the dephasing time of the wave-packet on the excited state. The dephasing time measured as a function of time-delay after pump excitation shows oscillations indicating oscillatory wave-packet dynamics on the excited state. From the experimental measurements and supporting theoretical calculations, we deduce that the wave-packet completely leaves the S$_1$ state surface after three traversals of the inter-system crossing between the singlet S$_1$ and triplet T$_2$ states.  
\end{abstract}

\maketitle

Ultrafast dynamics on excited electronic potential energy surfaces in molecules can occur on time scales ranging from a few femtoseconds to hundreds of picoseconds. In polyatomic molecules with multiple degrees of freedom, conical intersections exist between different electronic potential energy surfaces that allow efficient redistribution of population between different electronic states and, in some cases, rapid nonradiative relaxation to the ground electronic state\cite{worth_beyond_2004}. Such dynamics may be probed by femtosecond time-resolved techniques such as time-resolved photoelectron spectroscopy \cite{champenois_involvement_2016, champenois_ultrafast_2019} or transient absorption spectroscopy\cite{timmers_disentangling_2019}, both of which can measure the evolution of electronic binding energies on ultrafast timescales. In transient absorption spectroscopy, a small fraction of the probe light is absorbed by a sample that interacts with a pump pulse. Probe photons not absorbed by the pumped molecules constitute a background that must be subtracted to determine the change in absorption, which can limit sensitivity, particularly when performed with light sources having limited spectral or intensity stability such as ultrashort pulsed extreme ultraviolet or X-ray light sources. 

Nonlinear spectroscopies employing four-wave mixing (FWM) techniques such as Degenerate Four-Wave Mixing\cite{de_nalda_investigation_2014,ding_efficient_2013,marroux_multidimensional_2018} or Optical-Kerr Effect (OKE) \cite{mcmorrow_femtosecond_1988,palese_femtosecond_1994,zhu_optical_2005}, on the other hand, can be background-free. These FWM methods provide information on excited state dynamics when the pulses involved are resonant with the state of interest. However, they rely on electronic coherence between the ground and excited states that is established by the first pulse in the FWM sequence. This electronic coherence may not persist long enough to allow the excited state dynamics  to be followed completely, due to decoherence caused by inter-nuclear motion. Thus, in order to comprehensively measure excited state dynamics, measuring the non-linear response of the excited molecule is necessary. This can be achieved by introducing an additional excitation pulse that first populates specific electronic states, which are probed by four-wave mixing. In this work we demonstrate that a nonlinear probe, combined with an electronic excitation pump pulse, can discriminate between different electronic states, even near inter-system crossing geometries, where the potential energy surfaces are degenerate. 

The electronic excited states of nitrobenzene have been previously investigated experimentally\cite{hause_roaming-mediated_2011,blackshaw_nonstatistical_2019,takezaki_relaxation_1998,schalk_time-resolved_2018} and theoretically\cite{xu_computational_2005,takezaki_geometries_1997,mewes_molecular_2014,giussani_insights_2017}. Transient grating spectroscopy experiments\cite{takezaki_nonradiative_1997,takezaki_relaxation_1998} measured three different time scales for dynamics after excitation of the singlet S$_1$ state. The shortest time-scale measured in that study was $\sim$100~fs for the internal relaxation dynamics on the S$_1$ state before the wave-packet reaches an inter-system crossing (ISC) with the triplet T$_2$ state. A more recent theoretical study\cite{giussani_insights_2017} provides two explanations for the rapid decay times in the transient grating experiments\cite{takezaki_relaxation_1998}. The $\sim$100~fs lifetime is attributed to either the time taken to reach the S$_1$/T$_2$ ISC, or the time to reach a conical intersection between the excited S$_1$ and ground S$_0$ electronic states. A very high (0.8) quantum yield\cite{takezaki_nonradiative_1997} of the triplet T$_2$ is consistent with the calculated strong coupling between S$_1$ and T$_2$ states\cite{giussani_insights_2017} at the ISC. This strong coupling indicates that the ISC between S$_1$ and T$_2$ is the dominant pathway responsible for the decay of S$_1$ in $\sim$100~fs observed in the transient grating experiments\cite{takezaki_relaxation_1998}. Here, we measure the dephasing time of the excited wave-packet as a function of time delay after excitation. From this measurement we deduce that the wave-packet exhibits oscillatory dynamics on the S$_1$ state, with a period of $\sim 200$~fs, and completely leaves the S$_1$ state after three traversals of the ISC.   


\begin{figure}[t]
\includegraphics[width=0.75 \textwidth]{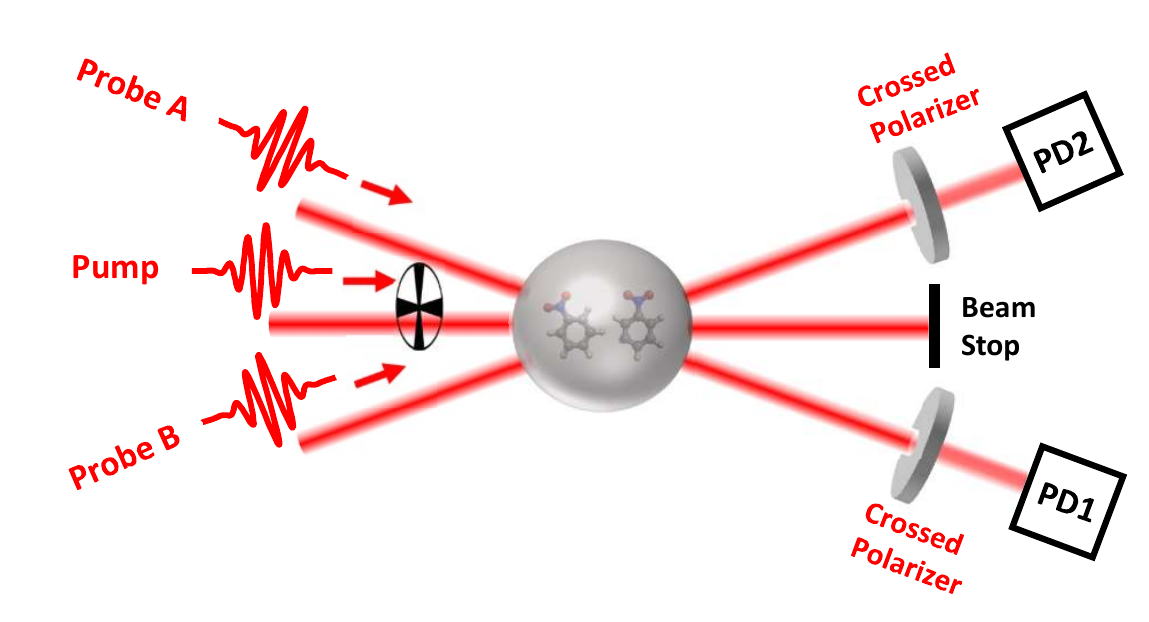}
\caption{\label{fig:1} A schematic of the UTPS experimental setup. The pump, probe $A$, and probe $B$ pulses are spatially and temporally overlapped at the nitrobenzene target. The polarization of probe $A$ is fixed at 45$^o$ with respect to the probe $B$ polarization. The pump pulse polarization can be varied using a half-wave plate. After interacting with the target, the probe $A$ and $B$ beams go through ultra-high contrast polarizers and are focused on photo-diode PD1 and PD2, respectively. The pump beam is blocked by a beam stop. Lenses used to focus the beams onto photo-diodes are not shown.}
\end{figure}

We measure the third-order non-linear response of liquid nitrobenzene molecules that are first excited to the S$_1$ electronic state by two-photon absorption of a pump pulse. In addition to the pump pulse, we use two pulses that together form a probing pulse pair. We use the Optical-Kerr Effect to measure the non-linear response of the pump-excited molecules. We call this new technique Ultrafast Transient Polarization Spectroscopy (UTPS). This scheme is similar to the method of pump-Degenerate Four-Wave Mixing (pump-DFWM) \cite{marek_direct_2011} but offers several advantages. Firstly, UTPS uses only two pulses in the probing sequence as opposed to three in pump-DFWM. Secondly, the simple phase matching conditions of the OKE sends the signal along the same direction as one of the probing pulses, which simplifies the experimental set-up. Finally, since the signal is emitted along the direction of either probing pulse, two signals can be measured simultaneously to obtain additional information in one time delay-scanning experiment. The relative simplicity of this scheme to measure excited state third-order non-linear response allows for the extension of the pump wavelength to vacuum-ultraviolet (VUV), extreme-ultraviolet (XUV) and even soft x-ray regimes by using a high-order harmonic generation or a free electron laser source. 

The experimental setup is shown in Figure~\ref{fig:1}. Near-infrared (IR) pulses with a central wavelength of 780 nm and pulse duration of $\sim 45$ fs are first split into two beams using a 50/50 beam-splitter. One arm forms the pump beam and the other is split again with a second 50/50 beam-splitter into two arms creating beams for probe pulses $A$ and $B$. In this measurement, pulses $A$ and $B$ together play the role of a probe. The time delay ($T$) between the pump and probe $A$ pulses, as well as the delay between the $A$ and $B$ probe pulses ($\tau$) are varied using two separate optical delay stages. When $T$ is varied at fixed $\tau$, the delay of both probe pulses A and B with respect to the pump varies at the same time.  All three pulses intersect at the sample target in a small angle non-collinear geometry. The time-smearing introduced by the crossing angle is small compared to each pulse duration. The spatial and temporal overlap between the three femtosecond pulses is found pair-wise, using second harmonic generation in a beta barium borate (BBO) crystal. The experimental target consists of a thin Spectrosil quartz cuvette with a wall thickness of 1~mm and a sample path length of 1~mm, filled with liquid nitrobenzene. After passing through the sample, the pump beam is blocked by a beam stop. The probe $A$ beam passes through an ultra-high contrast polarizer (extinction ratio $> 10^{6}$) and is focused on to photo-diode 1 (PD1) using a lens. The signal measured by PD1 is the UTPS signal of the probe $A$ beam. Similarly, the UTPS signal of probe $B$ is measured by photo-diode 2 (PD2). The intensity of the pump pulse is $\sim 1 \times 10^{11}$ W/cm$^2$ and the intensity of the probing pulses is $\sim 3 \times 10^{10}$ W/cm$^2$. The pump beam is modulated using a chopper wheel, which provides a reference for two lock-in amplifiers. The signal from each detector PD1 and PD2 is sent to a separate lock-in amplifier, and the output of each lock-in amplifier is then recorded using a data acquisition device and a desktop computer. 

The UTPS polarization signal measured in this optical homodyne configuration is related to the effective $\chi^{(3)}$\cite{palese_femtosecond_1994,lotshaw_intermolecular_1995}, henceforth denoted by $\chi^{(3)}_{eff}$,  as a function of the time delays between the pulses. The measured UTPS signal at PD1 can be written as

\begin{equation}
\begin{aligned}
\label{eqn:1}
I_{raw}(\tau, T) &= \int{\Big|E_{sig,A}(\omega,\tau, T)\Big|^2}d\omega \\
&= \int\Big|\Big[\chi^{(3)}_{eff,g}(\omega, \tau)|E_{B}(\omega)|^2E_{A}(\omega)+ \chi^{(3)}_{eff,g}(\omega, T)|E_{pump}(\omega)|^2E_{A}(\omega) \\
+ \chi^{(3)}_{eff,ex}(\omega, \tau, T)|E_{B}(\omega)|^2E_{A}(\omega)\Big]\Big|^2 d\omega 
\phantom{\hspace{-6cm}}
\end{aligned}
\end{equation}

where $\chi^{(3)}_{eff,g}$ and $\chi^{(3)}_{eff,ex}$ are the effective third-order susceptibilities, due to the ground state and excited state populations, respectively. $E_{A}$, $E_{B}$, and $E_{pump}$ are the corresponding spectral amplitudes of the electric-fields of the two probing pulses and the pump pulse. The terms involving all three spectral amplitudes of the pump and the two probe pulses are not included, because the signal from such an interaction will not reach the detector due to phase matching constraints. Upon expanding Equation~\ref{eqn:1} above, we see that several cross terms, containing both $\chi^{(3)}_{eff,g}$ and $\chi^{(3)}_{eff,ex}$, contribute to the signal in addition to the individual ground state and excited state contributions. By using lock-in amplification with the modulation of the pump beam as reference and by separately measuring the UTPS signals with one of the relevant probe pulses blocked, the two ground state-only contributions can be removed (see Supporting Material). Equation~\ref{eqn:1} can be further simplified to show that the signal depends on the real and imaginary parts of $\chi^{(3)}_{eff,g}$ and $\chi^{(3)}_{eff,ex}$ (see equations S2 and S3 in the Supporting Material). Here, we focus on obtaining information about excited state dynamics through the dephasing times, extracted from the UTPS measurements. The $\tau$ and $T$ dependencies of the spectral amplitudes of the probe pulses appear as phase factors, which vanish upon taking the absolute magnitude and are not shown in Equation~\ref{eqn:1}.

\begin{figure*}[t]
\includegraphics[width=0.95 \textwidth]{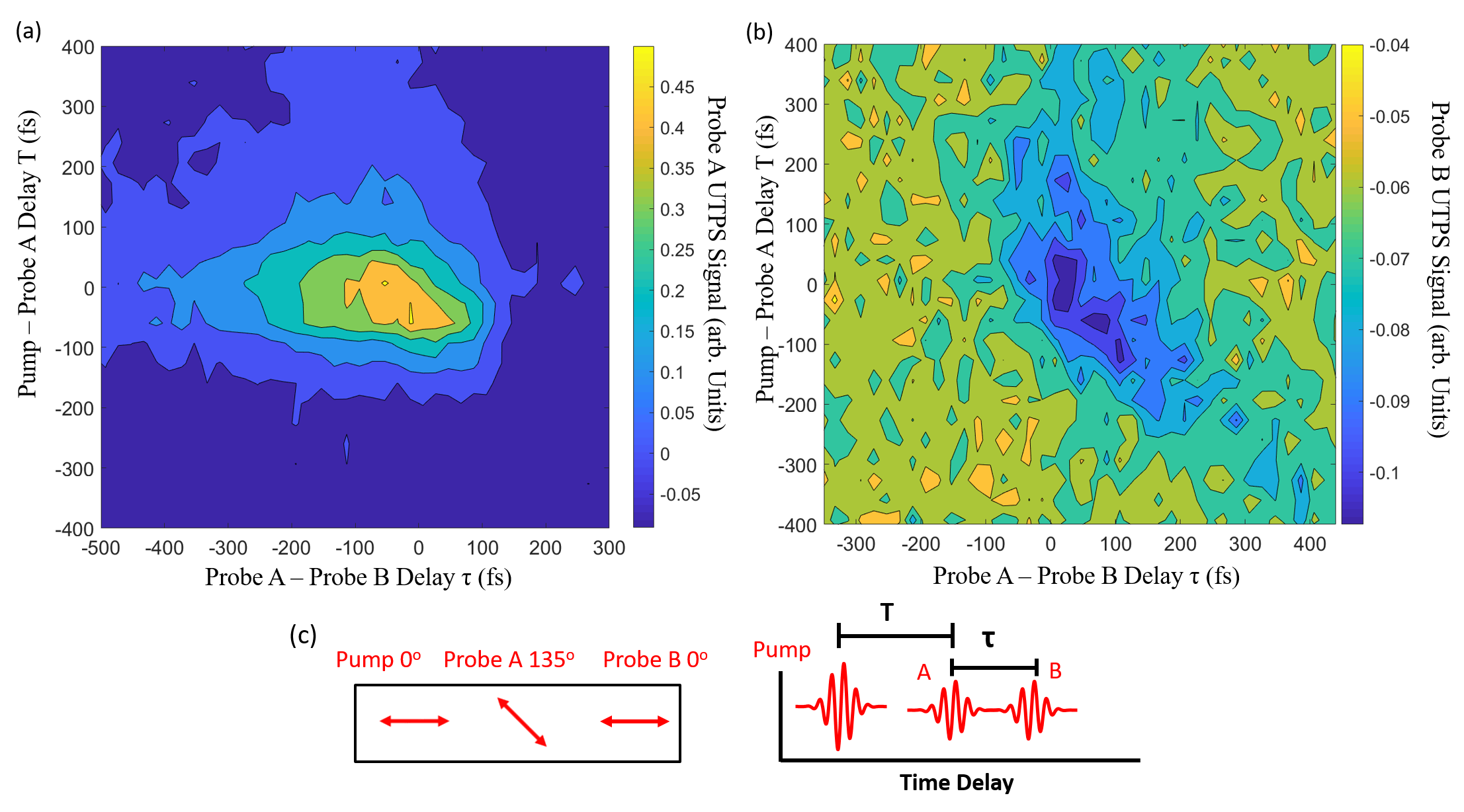}
\caption{\label{fig:2} The S$_1$ excited state ultrafast transient polarization signal measured for liquid nitrobenzene, using near infrared femtosecond pulses. (a) UTPS signal measured along the direction of probe pulse $A$ and (b) UTPS signal  measured along the direction of probe pulse B, both as a function of the two time-delays shown in (c). The 2-pulse contributions from the interaction of any two pulses have been subtracted. (c) Sketch of the polarization angles and time delays between the pump, probe $A$, and probe $B$. At positive $T$ values the pair of pulses $A$ and $B$ arrive after the pump. At positive $\tau$ values pulse $B$ arrives after pulse $A$.}
\end{figure*} 

The polarization signals of probe $A$ (PD1) and of probe $B$ (PD2) as a function of the time delays $\tau$ and $T$, after subtraction of the ground state-only contributions, are shown in Figures~\ref{fig:2}~(a) and (b) respectively. Figure~\ref{fig:2}~(c) indicates the polarization angles of the three pulses and the time delays between them. Each UTPS signal varies with both time delays $\tau$ and $T$. Below, we focus only on the probe $A$ UTPS signal. The probe $B$ UTPS signal, though acquired simultaneously, has significantly higher noise compared to that of probe $A$. The reason for the low signal in probe $B$ could be mainly due to imperfections in the spatial mode of probe $B$, which affects its spatial overlap with the other pulses. In addition to this, since probe B has a different polarization compared to probe A, the components of the third-order response that contribute to the signal may be weaker.

\begin{figure*}[t]
\includegraphics[width=0.8 \textwidth]{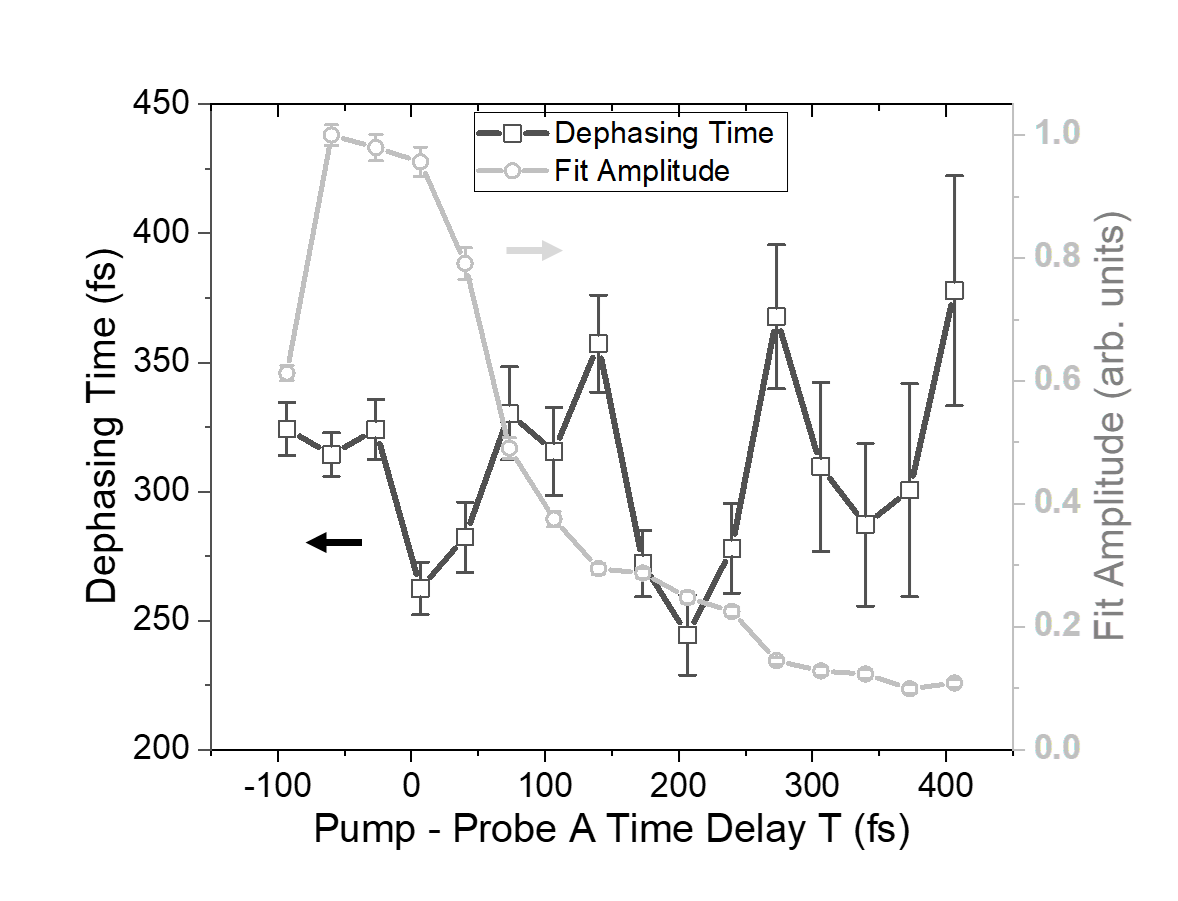}
\caption{\label{fig:3} Dephasing times $\gamma(T)$ (black squares and solid black line) for delay scans of the probe pulses $A$ and $B$  ($\tau$ axis), extracted using the fit Equation~\ref{eqn:2}, as a function of the pump-probe A delay $T$, for probe $A$ polarization signal (Figure~\ref{fig:2}~(a)). Error bars are one standard deviation and become very large beyond $T=300$ fs, as the signal vanishes at larger values of $T$. $\gamma(T)$ is a measurement of the dephasing time of vibrational coherences on the S$_1$ excited state and shows oscillations with a period of $\sim$200 fs. Grey circles and solid grey line show the fit amplitude $A(T)$ with error bars representing one standard deviation.}
\end{figure*}

 As shown in the Supporting Material, only $\chi^{(3)}_{eff,ex}$ is a function of both $\tau$ and $T$. In order to extract excited state information, we fit lineouts of the probe $A$ UTPS signal, taken at different values of $T$, with the following equation, which is a convolution of a Gaussian function with an exponential decay. A step function is also included to model any long decay dynamics that does not return the signal back to zero in the range of delay shown here:

\begin{equation}
\label{eqn:2}
\begin{split}
I_{sig}(\tau;T) = A(T)[H(\tau-t_0(T))exp(-\frac{\tau - t_0(T)}{\gamma(T)})]\circledast [g(\tau;\sigma)] 
\end{split}
\end{equation}

where $A(T)$ is the amplitude, $H(\tau-t_0(T))$ is the Heaviside step function, $t_0(T)$ is the onset time that quantifies the location of the rise of the signal, $\gamma(T)$ is the decay time of the exponential function, $g(\tau;\sigma)$ is a Gaussian function with width $\sigma$, which approximates the instrument response (the cross correlation function of the pulses).

The parameter $\gamma$ obtained from the fitting procedure, as a function of $T$, is plotted in Figure~\ref{fig:3} (black squares and solid black line). These decay times $\gamma(T)$ are a measure of the dephasing time of the excited state vibrational coherences. The dephasing time is seen to oscillate as a function of time $T$ after excitation of the molecule by the pump pulse. The oscillation amplitude is significantly larger than the error bars that represent one standard-deviation. Figure~\ref{fig:3} also shows the fit amplitude $A(T)$ with one standard deviation error bars. The fit amplitude decays and has a weak oscillation with a period of $\sim$ 200 fs. As shown in the Supporting Material, the measured signal has contributions from both the ground state and excited state third-order response, even after subtraction of the two ground state-only contributions in Equation~\ref{eqn:1}. However, only the excited state response $\chi^{(3)}_{eff,ex}(\tau, T)$ is dependent on both delay parameters $T$ and $\tau$. Thus the dephasing time $\gamma(T)$, measured for the $\tau$ axis as a function of $T$, should directly correspond to the excited state dynamics of S$_1$. On the other hand, $A(T)$ has contributions from both ground and excited states. As a result, the time scale of decay of the amplitude $A(T)$ does not correspond to the excited state lifetime. However, the oscillatory component of $A(T)$, which is in sync with the dephasing time oscillation, likely corresponds to the excited state dynamics.

\begin{figure}[t]
\includegraphics[width=0.75 \textwidth]{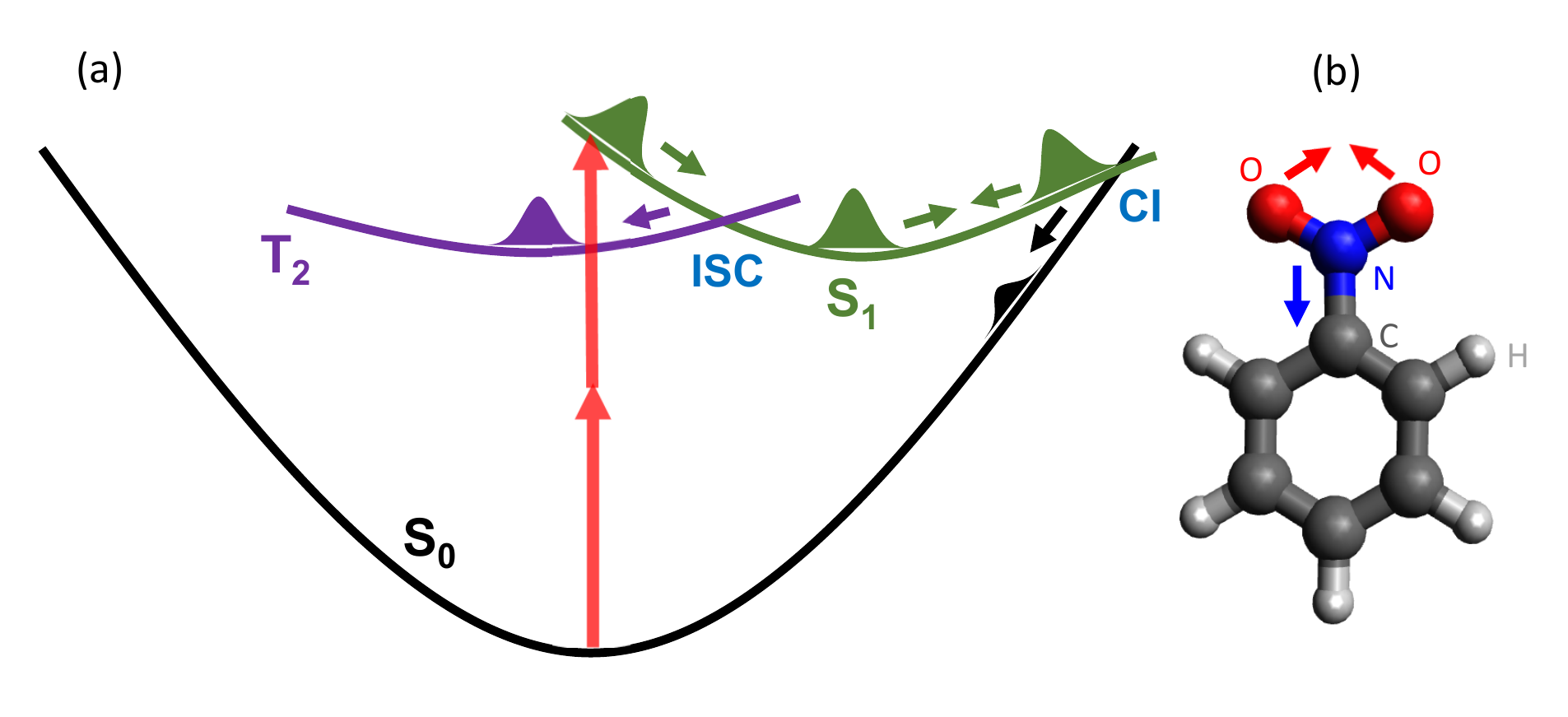}
\caption{\label{fig:4} (a) Schematic of the different pathways taken by the excited wave-packet based on reference \cite{giussani_insights_2017}. The wave-packet, excited by two photons from the pump pulse, moves towards the S$_1$/T$_2$ inter-system crossing (ISC). A portion of the wave-packet crosses to the T$_2$ state, and the remaining portion moves towards the S$_1$/S$_0$ conical intersection (CI). At the CI a small portion of the wave-packet crosses to the S$_0$ state, and the remaining portion returns towards the ISC to complete one full period of oscillation. The wave-packet then starts its second oscillation, and by the time it reaches the CI, most of the wave-packet leaves the S$_1$ state. (b) Structural illustration of nitrobenzene with red and blue arrows indicating O-N-O angle closing and C-N bond shortening, respectively, which are the dominant degrees of freedom as the wave-packet undergoes dynamics on the S$_1$ state\cite{giussani_insights_2017}.}
\end{figure}


 The oscillating $\gamma(T)$ is consistent with recent electronic structure calculations \cite{giussani_insights_2017}, which can be summarized as follows. After excitation to the S$_1$ state, the nuclear wave-packet undergoes significant O-N-O angle closing and C-N bond shortening. As illustrated in Figure~\ref{fig:4}, the wavepacket moves along the S$_1$ state surface and encounters the ISC region first\cite{giussani_insights_2017}. A portion of the wave-packet crosses here to the T$_2$ state. The remaining part of the wave-packet continues on the S$_1$ surface, primarily with further C-N shortening to 1.241~{\AA} and O-N-O angle closing to 94.77$^\circ$, to reach the S$_1$/S$_0$ CI, which is elevated by 0.6~eV\cite{mewes_molecular_2014} or 0.3~eV\cite{giussani_insights_2017} relative to the ISC potential energy. Since the coupling between the states at this CI is small\cite{giussani_insights_2017}, a small fraction of the wave-packet hops to the S$_0$ surface. The remaining S$_1$ wave-packet then returns on S$_1$ to the ISC region at $T\approx 200$~fs, after which only a small fraction of the wave-packet remains on S$_1$. After encountering the ISC region a third time, at $T\approx300$~fs, the small signal (Figure~\ref{fig:2}~(a)) and increasing error bars for the dephasing time (Figure~\ref{fig:3}) are consistent with a small fraction of the wavepacket remaining on the S$_1$ surface. Our measurements suggest that three encounters with the ISC region are necessary for the population to completely leave the initially excited state.
\begin{table}[t]
\caption{\label{tab:theory}
Geometry-dependence of single molecule $\chi^{(3)}_{eff}$ (arbitrary units) for the S$_0$, S$_1$, and T$_2$ electronic states, calculated using the sum-over-states method. All geometries are from Giussani and Worth\cite{giussani_insights_2017}.}
\begin{tabular}{lrrr}
Geometry &S$_0$ &S$_1$ &T$_2$\\
\hline
S$_0$ minimum &5.714 &123.967 &14.292\\
S$_1$ minimum &4.452 &8.375 &7.467\\
T$_2$ minimum &3.704 &9.852 &8.318\\
S$_1$/S$_0$ CI &7.514 &9.036 &7.863\\
S$_1$/T$_2$ ISC &4.487 &8.262 &7.207\\
\end{tabular}
\end{table}
This picture is supported by our theoretical calculations of $\chi^{(3)}_{eff}$ in the ground and excited states, which are summarized in Table~\ref{tab:theory} (see Supporting Material for details). From the calculations we expect the signal to be dominated by $\chi^{(3)}_{eff,ex}$ due to the S$_1$ state in the Franck-Condon region, with the S$_0$ and T$_2$ contributions being smaller by an order of magnitude or more. Likewise, S$_1$ contributions to $\chi^{(3)}_{eff}$ decrease markedly in the other geometries. Therefore, as the wave-packet evolves along the S$_1$ surface, the $\chi^{(3)}_{eff}$ signal is lost due to (i) nonadiabatic transitions to the S$_0$ and T$_2$ states, and (ii) nuclear wavepacket motion away from the Franck-Condon region.

In conclusion, we have demonstrated that ultrafast time-resolved measurements of the third-order nonlinear response of a photoexcited liquid is sensitive to nonadiabatic transitions between excited electronic states, providing information that can be used to track ultrafast molecular dynamics. Using the optical-Kerr effect along with a pump pulse in the Ultrafast Transient Polarization Spectroscopy scheme can offer significant advantages. The dephasing time of vibrational coherences on the excited S$_1$ state oscillate as a function of the time delay after pump excitation. This variation in dephasing time indicates oscillatory behavior of the excited wave-packet. We find that three encounters of the ISC region by the wave-packet are necessary for the population to completely leave the S$_1$ surface. The approach of UTPS has the potential for sensitively probing wave-packet dynamics near conical intersections in molecules, and the simple optical instrumentation is compatible with applications involving a wide range of photon energies, including extreme-ultraviolet and soft x-rays. Future experimental developments could combine third-order response measurements by UTPS with transient absorption spectroscopy, to measure binding energies and electronic character of the states involved, allowing ultrafast transient features to be revealed within congested spectra by multidimensional nonlinear spectroscopy.

\begin{acknowledgement}
This work was supported by the U.S. Department of Energy, Office of Science, Office of Basic Energy Sciences, Chemical Sciences, Geosciences, and Biosciences
Division. 
Work at the Molecular Foundry was supported by the Office of Science, Office of Basic Energy Sciences, of the U.S. Department of Energy under Contract No. DE-AC02-05CH11231. This research used resources of the National Energy Research Scientific Computing Center, a DOE Office of Science User Facility supported by the Office of Science of the U.S. Department of Energy under Contract No. DE-AC02-05CH11231.
\end{acknowledgement}

\bibliography{UTPSnitrobenzene.bib}

\end{document}